\documentclass[conference]{IEEEtran}
\IEEEoverridecommandlockouts
\usepackage{cite}
\usepackage{amsmath,amssymb,amsfonts}
\usepackage{graphicx}
\usepackage{textcomp}
\usepackage{cuted}
\usepackage[usenames,dvipsnames]{xcolor}

\usepackage[utf8]{inputenc} 
\usepackage[T1]{fontenc}    
\usepackage{hyperref}       
\usepackage{url}            
\usepackage{booktabs}       
\usepackage{nicefrac}       
\usepackage{microtype}      
\usepackage{lipsum}
\usepackage{float}
\usepackage{braket}
\usepackage{enumitem}

\usepackage[caption=false,font=footnotesize]{subfig}

\newcommand{\toronto}{toronto~}
\newcommand{\yorktown}{yorktown~}
\newcommand{\athens}{athens~}
\newcommand{\rochester}{rochester~}

\def\BibTeX{{\rm B\kern-.05em{\sc i\kern-.025em b}\kern-.08em
    T\kern-.1667em\lower.7ex\hbox{E}\kern-.125emX}}

\begin{document}

\title{Stability of Noisy Quantum Computing Devices}%

\author{Samudra~Dasgupta$^{1,2}$ and Travis~S.~Humble$^{1,2}$\\
\small{$^1$Quantum Science Center, Oak Ridge National Laboratory, Oak Ridge, Tennessee, USA}\\
\small{$^2$Bredesen Center, University of Tennessee, Knoxville, USA}\\
\thanks{This manuscript has been authored by UT-Battelle, LLC under Contract No. DE-AC05-00OR22725 with the U.S. Department of Energy. The United States Government retains and the publisher, by accepting the article for publication, acknowledges that the United States Government retains a non-exclusive, paid-up, irrevocable, worldwide license to publish or reproduce the published form of this manuscript, or allow others to do so, for United States Government purposes. The Department of Energy will provide public access to these results of federally sponsored research in accordance with the DOE Public Access Plan. (http://energy.gov/downloads/doe-public-279access-plan).}
}


\maketitle
\thispagestyle{plain}
\pagestyle{plain}

\begin{abstract}
Noisy, intermediate-scale quantum (NISQ) computing devices offer opportunities to test the principles of quantum computing but are prone to errors arising from various sources of noise. Fluctuations in the noise itself lead to unstable devices that undermine the reproducibility of NISQ results. Here we characterize the reliability of NISQ devices by quantifying the stability of essential performance metrics. Using the Hellinger distance, we quantify the similarity between experimental characterizations of several NISQ devices by comparing gate fidelities, duty cycles, and register addressability across temporal and spatial scales. Our observations collected over 22 
 months reveal large fluctuations in each metric that underscore the limited scales on which current NISQ devices may be considered reliable.
\end{abstract}

\section*{Introduction}
On-going efforts to realize the principles of quantum computing have demonstrated high-fidelity control over quantum physical systems ranging from superconducting electronics \cite{burnett2019decoherence}, trapped ions \cite{wan2019quantum}, and silicon \cite{PhysRevApplied.10.044017} among many others \cite{humble2019quantum}. As these efforts aim for future fault-tolerant operation \cite{gottesman1998theory}, they currently establish noisy, intermediate-scale quantum (NISQ) devices as a frontier for testing quantum computing under experimental conditions \cite{preskill2019quantum}. As first-in-kind platforms, NISQ computing devices enable design verification \cite{PhysRevA.101.042315}, device characterization \cite{harper2020efficient}, program validation \cite{ferracin2019accrediting}, and a breadth of testing and evaluation for application performance \cite{kandala2017hardware,dumitrescu2018cloud,hempel2018quantum,klco2018quantum,mccaskey2019quantum,roggero2020quantum} with several recent demonstrations exemplifying the milestone of quantum computational advantage \cite{arute2019,google2020hartree,zhong2020quantum}.
\begin{figure}[htbp]
\begin{center}
\subfloat[]{%
  \includegraphics[width=0.4\columnwidth]{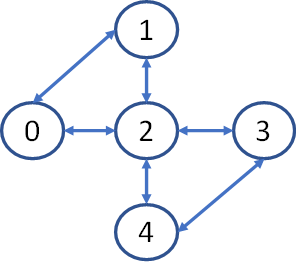}%
  \label{fig:yorktown}
}

\subfloat[]{%
  \includegraphics[width=\columnwidth]{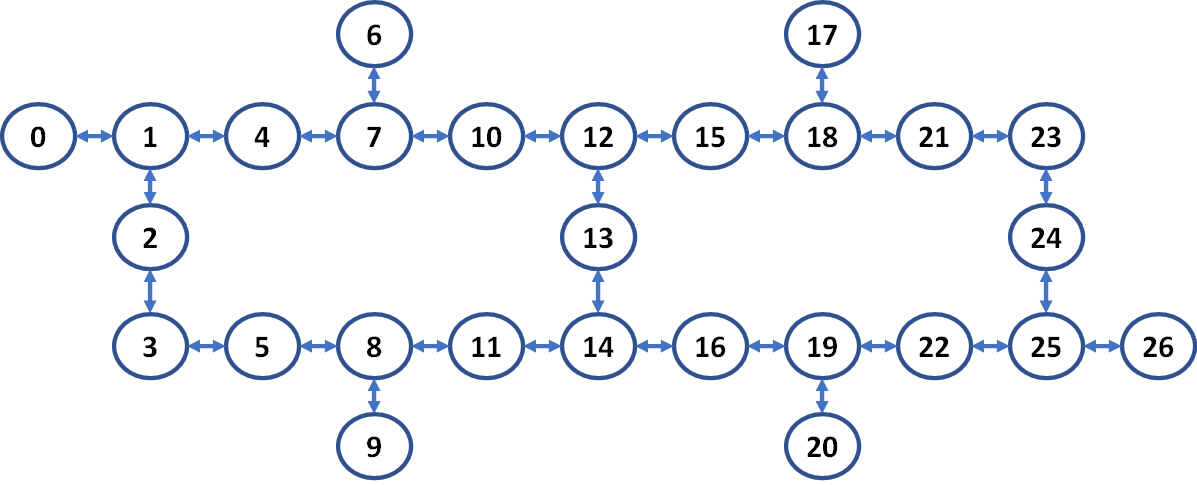}%
  \label{fig:toronto}
}
\caption{
Schematic layout of the \protect\subref{fig:yorktown} \yorktown and \protect\subref{fig:toronto} \toronto devices produced by IBM. Circles denote register elements and edges denote connectivity of 2-qubit operations.
}
\label{fig:yorktown_and_toronto}
\end{center}
\end{figure}
\par
A prominent feature of experimental NISQ computing is that noise and errors limit the behavior of these devices \cite{martinis2015qubit}. Characterization methods to quantify noise in NISQ devices inform benchmarking methods that measure the resulting quantum computational state \cite{eisert2020quantum,BlumeKohout2020volumetricframework,lilly2020modeling}. Reliable benchmarks establish bounds on the statistical significance of the observed results and help clarify the conditions necessary for results to be reproducible. However, experimental characterizations of current NISQ devices reveal transient changes in noise are prominent \cite{proctor2020detecting}. Even when temporal drift and spatial variability are mitigated efficiently on short scales, e.g., the duration of a single program execution \cite{temme2017error,kandala2019error,geller2020rigorous,wilson2020just,hamilton2020error}, unstable noise undermines efforts to track progress in device performance and reproduce experimental milestones.
\par
Here we address the stability of NISQ devices as a key concern in the reproducibility of quantum computing benchmarks. We define stability relative to fluctuations in noisy behavior across space and time, where a stable device is defined to exhibit noise that is consistent across these different scales within a desired tolerance. We present a suite of metrics by which to evaluate stability in current NISQ devices, and we experimentally quantify stability of several example devices using statistical tests. Device stability is directly related to the reproducibility of experiments in NISQ computing, and our analysis identifies when a NISQ device is reliable and the experimental conditions under which benchmark results are reproducible.
\par
\begin{table}[htp]
\caption{Device metrics for assessing DiVincenzo criteria}
\label{tab:dvz}
\centering
\begin{tabular}{|p{3.0cm}|p{5.0cm}|}
    \hline 
    \textit{Metric} (\textit{Symbol}) & \textit{Description} \\ \hline
    Capacity ($n$) & Maximal amount of information that may be stored in the register.  \\ \hline
    {\raggedright Initialization Fidelity ($F_{\textrm{I}}$)} & Accuracy with which a fiducial register state is prepared. \\ \hline
    {\raggedright Gate Fidelity ($F_{\textrm{G}}$) } & Accuracy with which a gate transforms the register state. \\ \hline
    Duty Cycle ($\tau$) & Ratio of gate duration to coherence time. \\ \hline
    Addressability ($F_{\textrm{A}}$) & Mutual information between register elements. \\ \hline
    \end{tabular}
\end{table}
\par 
Our approach borrows criteria originally proposed by DiVincenzo as minimal for the realization of quantum computing \cite{divincenzo2000physical}. Table~\ref{tab:dvz} presents each criterion as quantified by a corresponding metric alongside a representative symbol and definition. A variety of methods exist for estimating each metric, and we select specific approaches to demonstrate the role of stability in observed results. For example, we characterize the capacity $n$ of the quantum register as the number of elements that can be addressed. We summarize approaches for the other metrics below (additional details can be found in Appendix A).
\par 
Initialization fidelity $F_I$ quantifies the accuracy with which a target quantum state is prepared in the register, and we quantify this fidelity in terms of the observed error following readout. Sophisticated tomographic approaches are generally more informative, but we find that characterization in a single computational basis state, $|0\rangle^{\otimes n}$, yields insights into the transient error rate $e_{R}$. The initialization fidelity itself is then defined as
\begin{equation}
F_I = 1 - e_R
\end{equation}
Similarly, gate fidelity $F_G$ measures the accuracy with which a quantum operation transforms the register state. We rely on randomized benchmarking of the 2-qubit Clifford group to track the error behavior of the CNOT operation \cite{aleksandrowicz2019qiskit}. This technique measures survival probability following a sequence of randomly selected Clifford elements and fits the resulting sequence of fidelity to a linear model that eliminates the influence of state preparation and measurement errors \cite{magesan2011scalable}. From the fitted parameters, the error per Clifford gate $\epsilon_G$ defines the given gate fidelity as
\begin{equation}
F_G = 1 - \epsilon_{G}
\label{eqn:fg}
\end{equation}
\par
The duty cycle $\tau$ is defined as the ratio of register coherence time to gate duration. Whereas gate duration represents the amount of time to complete the intended transform, the coherence time represents the timescale over which the encoded quantum information decoheres \cite{gambetta2019}. Using a simplified model of exponential decay that neglects other processes, decoherence of a single-qubit is characterized by a time-dependent decay with the characteristic time $T_2$. Hence,  the duty cycle is defined as 
\begin{equation}
    \tau = T_2/T_G
\end{equation}
with $T_G$ the duration for gate $G$.
\par
Finally, the addressability $F_A$ quantifies the ability to measure register elements individually. We define addressability in terms of pair-wise correlations that arise during measurement (in the computational basis) and quantify this metric using the normalized mutual information $\eta$. Given the entropy  $H(X)$ of a discrete random variable $X$, and joint entropy $H(X,Y)$ of random variables $X$ and $Y$, we define
the mutual information $I(X, Y)$ between two random variables $X$ and $Y$ as
\begin{equation}
I(X, Y) = H(X) + H(Y) - H(X,Y)
\end{equation}
The normalized mutual information $\eta$ is then obtained by normalizing the mutual information $I(X, Y)$ with the average entropy $H_{avg}(X,Y) = (H(X)+H(Y))/2$ such that
\begin{equation}
\eta(X,Y) = \frac{I(X, Y)}{H_{avg}(X,Y)}
\end{equation}
This yields the addressability
\begin{equation}
F_A = 1-\eta
\label{eq:FM}
\end{equation}
which is dependent on the measured state.  In particular, the addressability is unity for separable states and vanishes for maximally entangled states due to correlated outcomes.
\par
\par
We next evaluate the stability of these metrics using statistical tests to measure variations in the observed distributions for NISQ devices.
We use the Hellinger distance 
\begin{equation}
d_{H}(P_i,P_j) = \sqrt{1-BC(P_i,P_j)}
\end{equation}
as a similarity measure between the observed $i$-th and $j$-th distributions, $P_i$ and $P_j$, for a given metric, with the Bhattacharyya coefficient
\begin{eqnarray}
BC(P_i,P_j) &=&
\begin{cases}
    \int\limits_{x} \sqrt{p_i(x) p_j(x)} dx            &\text{(continuous)}\\
    \sum\limits_{x \in X} \sqrt{p_i(x) p_j(x)}         &\text{(discrete)}
\end{cases}
\end{eqnarray}
bounded between 0 and 1. The Hellinger distance itself vanishes for identical distributions and approaches unity for distributions with no overlap. 
\par 
The Hellinger distance provides insight into the fluctuations between the device metrics as well as the underlying quantum states. For example, the Hellinger distance between the measurement outcomes of a quantum circuit provides a lower bound on a similar distance between the underlying quantum states $\rho$ and $\sigma$   \cite{jarzyna2020geometric}. These distances coincide when the optimally discriminating measurement is the computational basis, such that we characterize the reproducibility of a program in terms of the Hellinger distance between computational outcomes. Variations in the device metrics describe how the corresponding quantum channels preparing those states fluctuate.

\begin{figure*}[t]
\subfloat[]{%
  \includegraphics[width=\columnwidth]{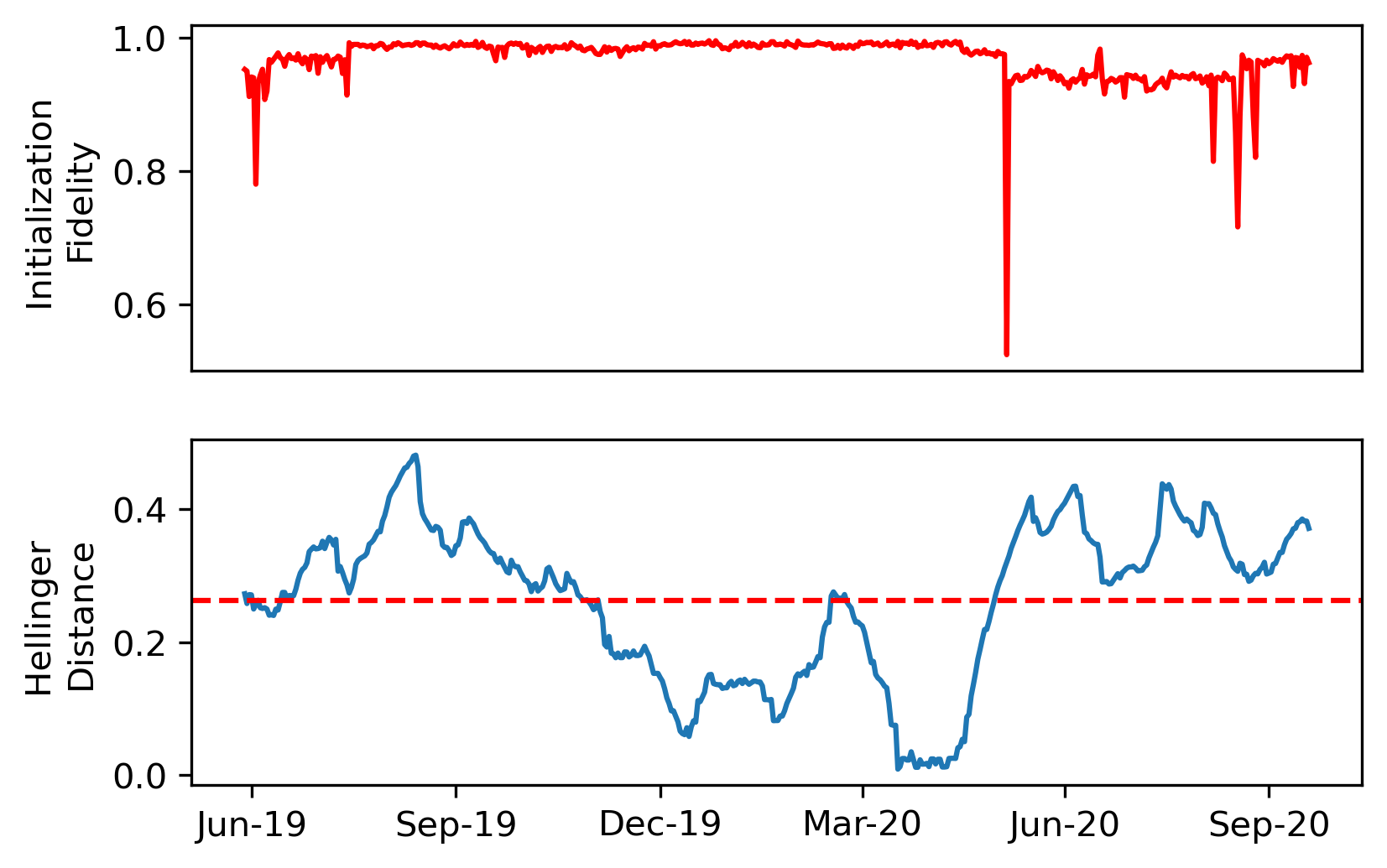}%
  \label{fig:FI_temporal_MBD_willow}
}
\hfill
\subfloat[]{%
\includegraphics[width=\columnwidth]{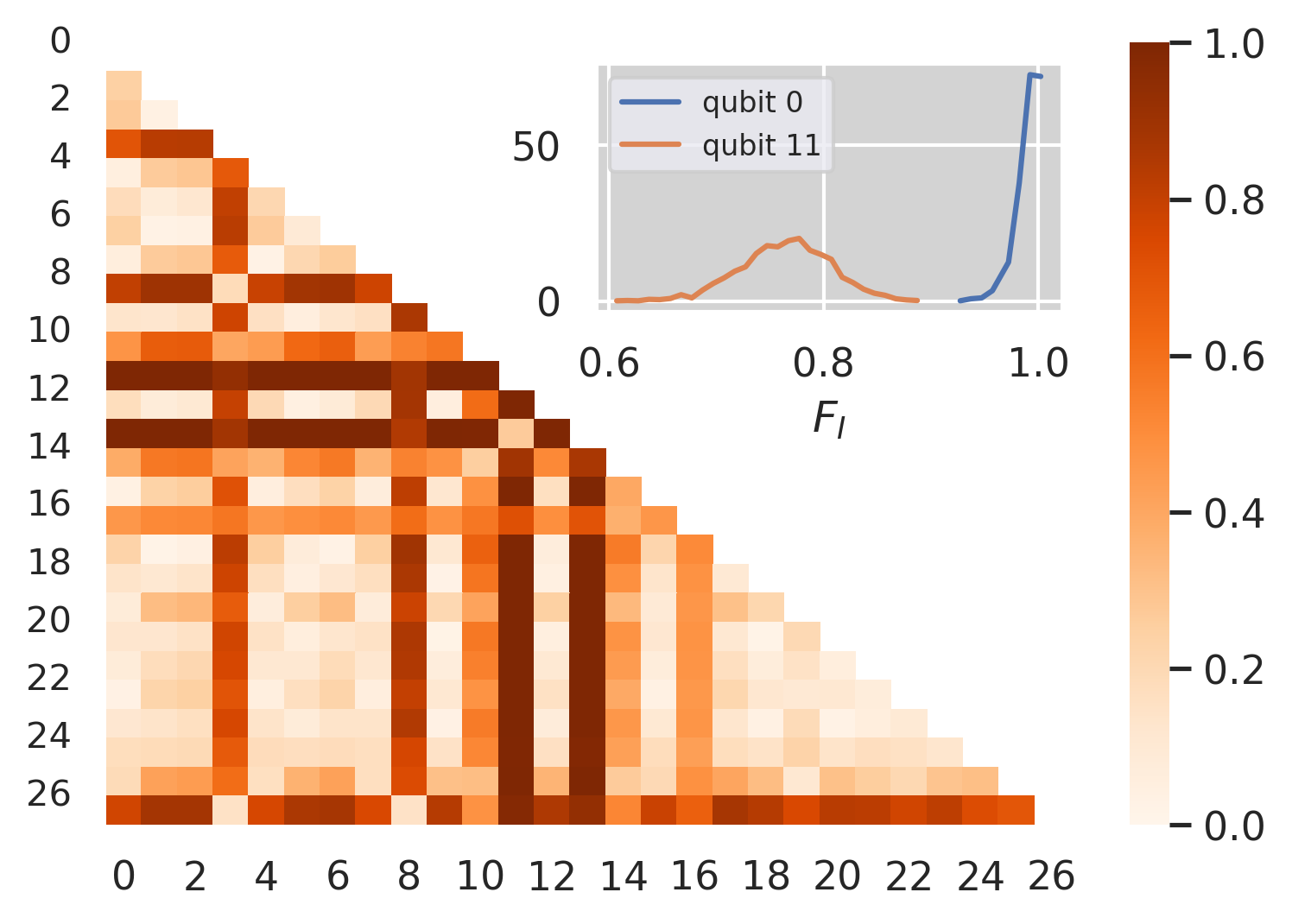}%
\label{fig:toronto_FI_spatial_birch}
}
\caption{
\protect\subref{fig:FI_temporal_MBD_willow} Temporal stability of the initialization fidelity $F_I$ for the \yorktown device. The top panel shows the average $F_I$ of register 0, and the bottom panel shows a running calculation of the Hellinger distance using a one-month window. The dashed red line is the median value.
\protect\subref{fig:toronto_FI_spatial_birch} Spatial stability of the initialization fidelity $F_I$ for the \toronto device sampled during 8:00-8:30 AM (UTC-5) on 11 December 2020. Hellinger distance between histograms for register pairs is measured. Inset show the distributions of $F_I$ for register 0 and 11, which have distance 1.0.
}
\label{fig:willow}
\end{figure*}
\section*{Results}
We evaluate the device metrics of Table~\ref{tab:dvz} with respect to temporal stability and spatial stability. Temporal stability evaluates changes in the distance between observed metrics at different times, while spatial stability evaluates how metrics vary across locations in the quantum register. We characterize in detail the stability of two superconducting transmon devices produced by IBM called \yorktown and toronto. With the layouts shown in Fig.~\ref{fig:yorktown_and_toronto}, these NISQ devices are leading examples of programmable platforms used for experimental validation of quantum computing applications. A detailed historical record quantifying the calibration of the $n=5$ \yorktown is accessible using the Qiskit toolset \cite{aleksandrowicz2019qiskit}. We collected daily calibration metrics of \yorktown from 1 March 2019 to 30 December 2020, and we collected a second data set from the $n=27$ \toronto device to characterize short-term and long-length scale stability using the addressability metric. The second dataset was collected directly from \toronto on 11 December 2020 during periods of time 8:00-8:00am, 11:00-11:00am, 2:00-2:30pm, 5:00-5:30pm, 8:00-8:30pm, and 11:00-11:30pm (UTC-5).
\subsection*{Initialization Fidelity}
We first evaluate temporal stability of the initialization fidelity for the \yorktown device. Figure~\ref{fig:FI_temporal_MBD_willow} shows the Hellinger distance over successive quarters for the initialization fidelity of register 0 from March 2019 to Dec 2020. Median distance plotted as a dashed red line emphasizes the aperiodic behavior in $F_I$ during this time. The inset shows the corresponding time series for $F_I$. While there are long spans during which $F_I$ itself is tightly controlled, the changes in the distance reveal the presence of large fluctuations in the underlying distributions. Similar behavior is observed for other registers in this device.
\par 
We also evaluate the spatial stability of the initialization fidelity. Here we use the $n=27$ \toronto device characterized at periodic intervals on a single day.  Figure~\ref{fig:toronto_FI_spatial_birch} plots the Hellinger distance for $F_I$ between pairs of register elements, showing large differences between the distributions observed at different locations.  While many distributions of the initialization fidelity within the \toronto device are similar, others are well separated as indicated by the darker bands. The inset plots histograms for registers 0 and 11, which have a Hellinger distance of 1.0.
\begin{figure*}[ht]
\subfloat[]{%
  \includegraphics[clip,width=\columnwidth]{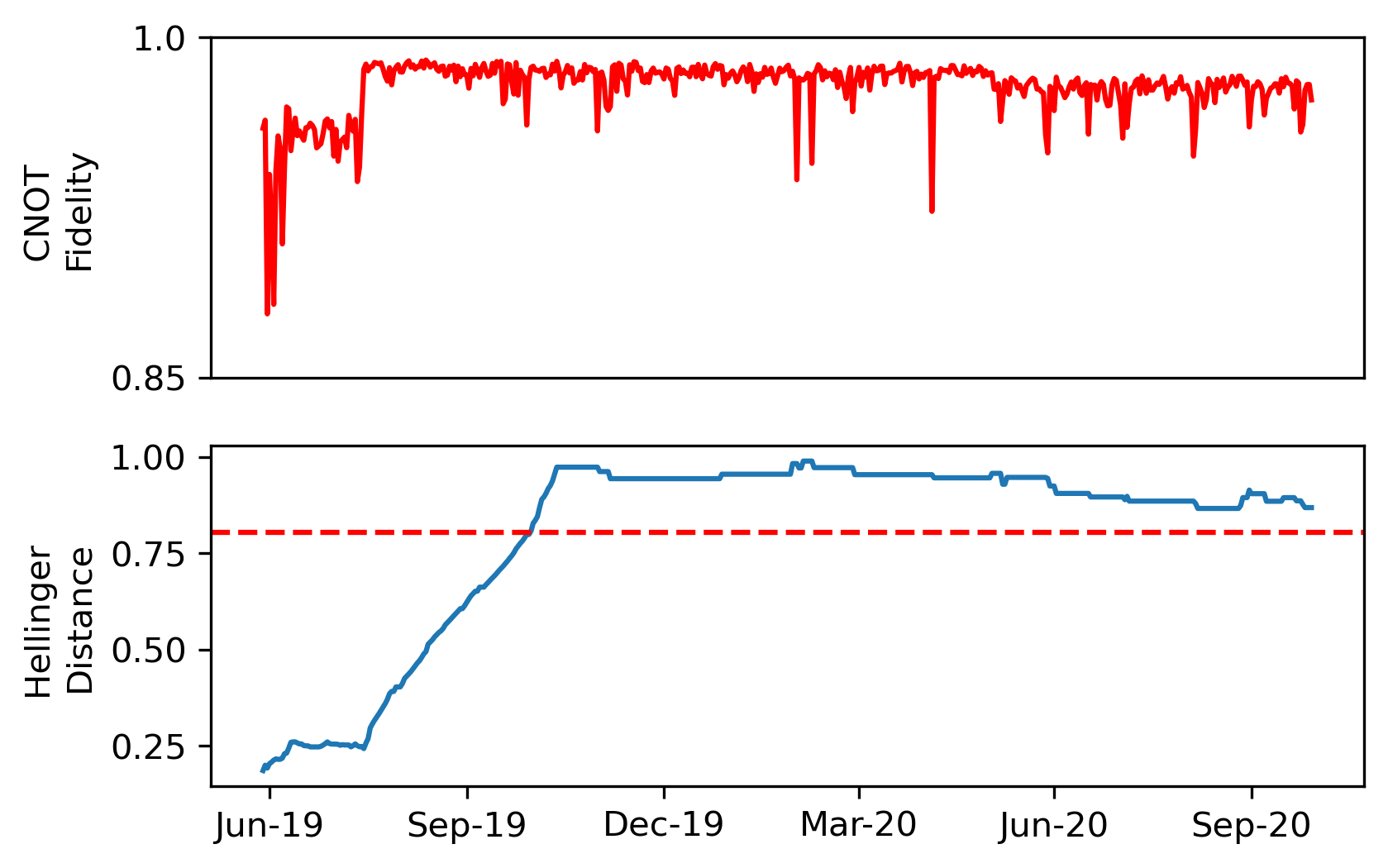}%
  \label{fig:yorktown_FG_temporal_maple}
}\hfill
\subfloat[]{%
  \includegraphics[clip,width=\columnwidth]{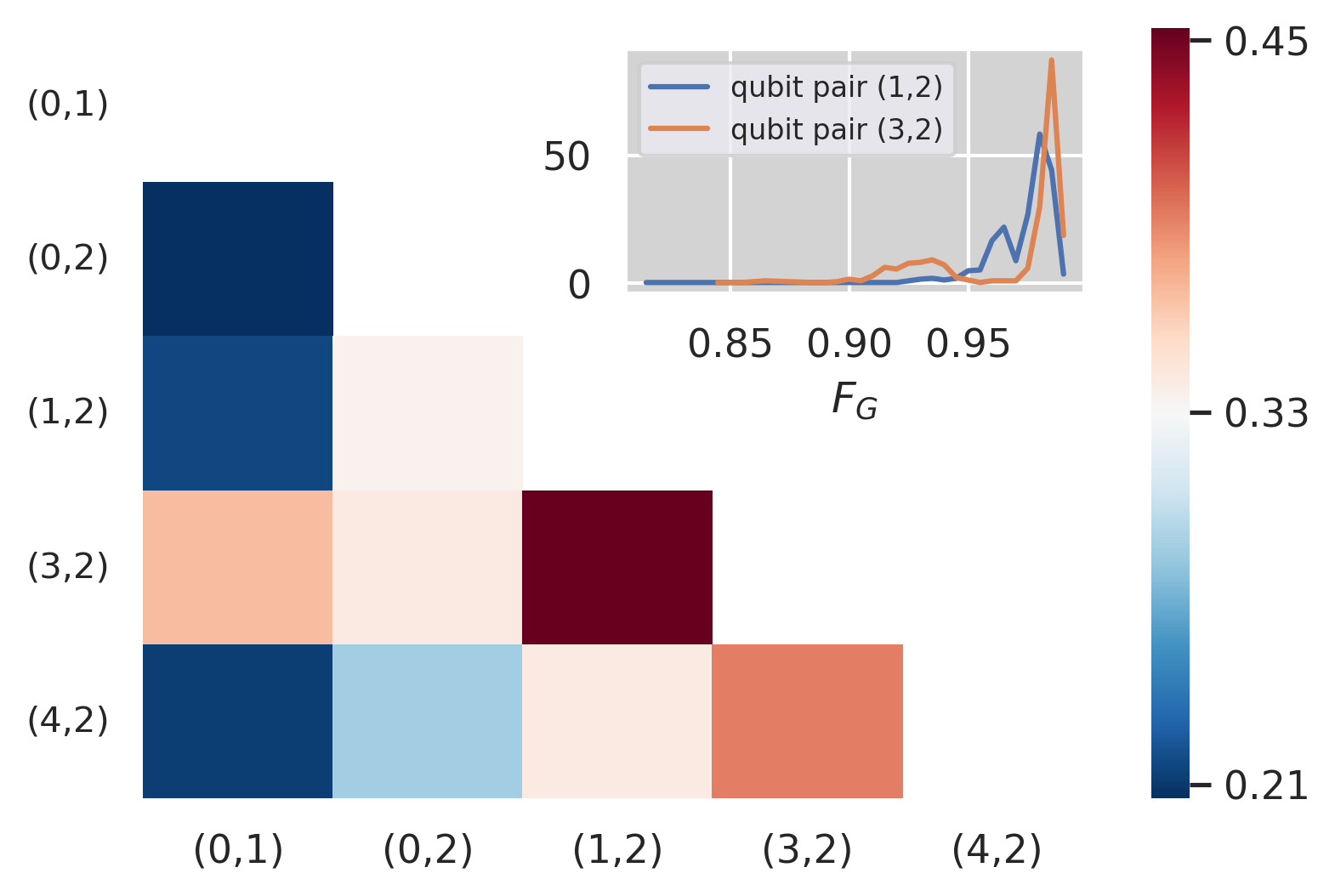}%
  \label{fig:yorktown_FG_spatial_fir}
}
\caption{
\protect\subref{fig:yorktown_FG_temporal_maple} 
Temporal stability of the gate fidelity $F_G$ for the CNOT gate for register pair (0,1) of the \yorktown device from March 2019 to December 2020. The top panel shows the average $F_G$ of register pair (0,1), and the bottom panel shows a running calculation of the Hellinger distance with respect to May 2019. The dashed red line is the median value.
\protect\subref{fig:yorktown_FG_spatial_fir} Spatial stability of the gate fidelity $F_G$ for the CNOT gate of the \yorktown device from March 2019 to December 2020. Inset shows the histograms 
with Hellinger distance 0.467.
}
\label{fig:fir}
\end{figure*}
\subsection*{Gate Fidelity}
We next analyze stability of gate fidelity $F_G$ for the CNOT gate in the \yorktown device. The error per gate $\epsilon$ was retrieved from March 2019 to December 2020, and  Fig.~\ref{fig:yorktown_FG_temporal_maple} plots the Hellinger distance for the corresponding CNOT fidelity of register pair (0,1). The stability of the fidelity is referenced to the initial distribution for $F_I$ collected in March 2019, and the metric diverged sharply during the period June 2019 to December 2019. However, the metric $F_G$ for register pair (0,1) fluctuated much less during the next 12 months. Whether this later behavior is sufficiently stable depends on the given application. Similar temporal analyses may be extended to all CNOT gates to generate a a detailed temporal map of the error model. However, it is clear from Fig.~\ref{fig:yorktown_FG_temporal_maple} that the noisy CNOT operations performed on March 2019 are vastly different from those performed in September 2020.
\par 
The spatial stability of $F_G$ for the \yorktown device is also shown in Fig.~\ref{fig:yorktown_FG_spatial_fir}. The heatmap measures the distance between the fidelities describing pairs of CNOT operations. It corresponds to the metric distribution for the entire data period, though temporal subsets could be easily generated. The greatest distance was found between pairs (1,2) and (3,2), and the presence of register 3 generally correlated with larger distances. The inset shows the probability distributions for two register pairs, (1,2) and (3,2), which yielded a distance of 0.467. This comparison highlights that the distributions may be peaked near similar values and yet still starkly dissimilar.
\subsection*{Duty Cycle}
\begin{figure}[htbp]
\begin{center}
\includegraphics[clip,width=\columnwidth]{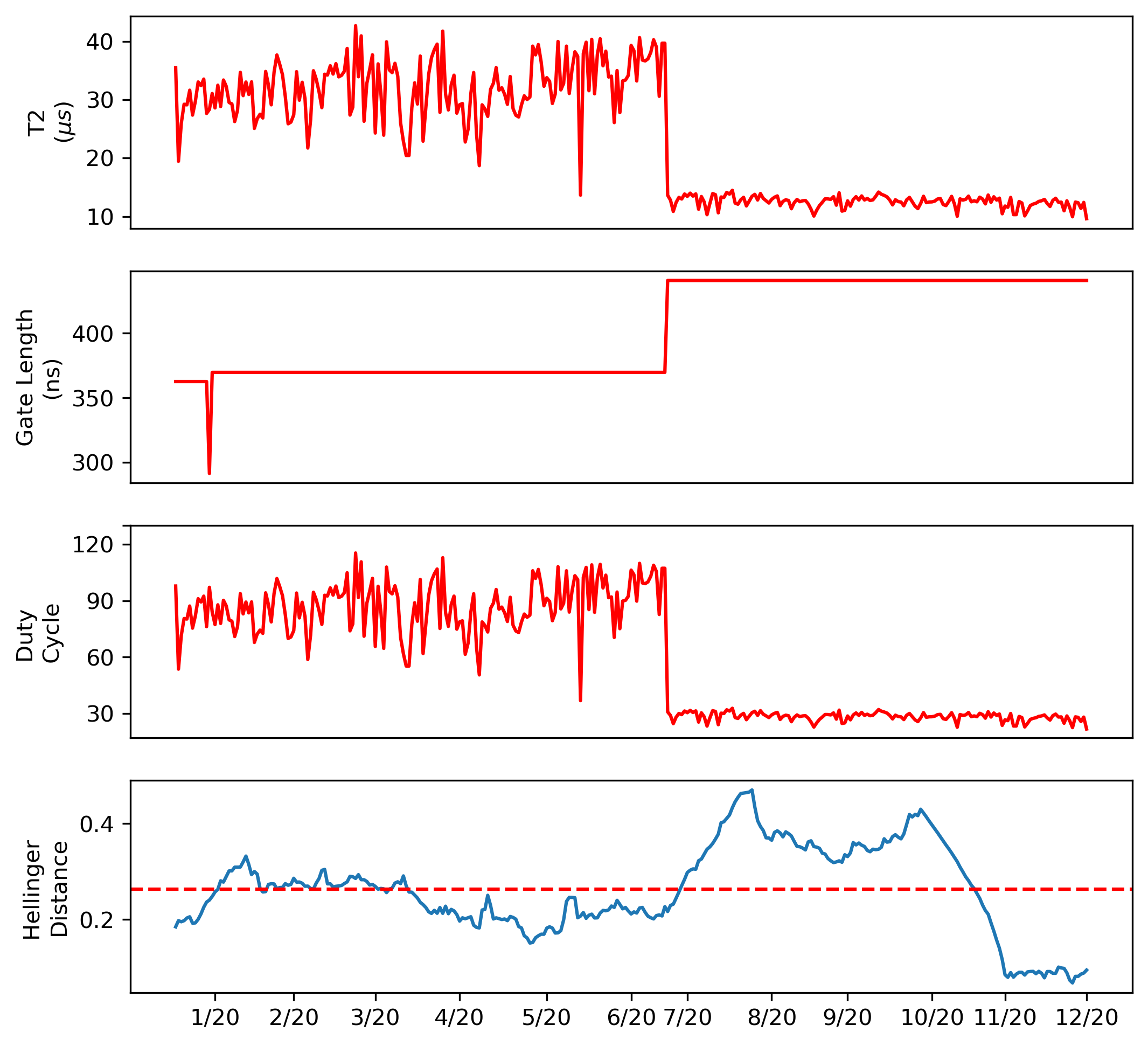}%
\caption{\protect Temporal stability of the CNOT duty cycle for register pair (0,1) in the \yorktown device. The top panel shows the harmonic mean of the register decoherence time $T_2$ for elements 0 and 1. The upper-middle panel shows the gate duration $T_G$ for the CNOT operation between these register elements. The lower-middle panel plots the corresponding duty cycle $\tau$ defined as the ratio of register decoherence time to gate duration. The bottom panel presents a running calculation of the Hellinger distance for the duty cycle distributions using a one-month window. The dashed red line is the median value.
}
\label{fig:yorktown_tau_temporal}
\end{center}
\end{figure}

Duty cycle monitors the ratio of coherence time to gate duration to yield a composite metric that estimates the number of operations that can be performed reliably. Figure~\ref{fig:yorktown_tau_temporal} presents the time series of these different metrics for a CNOT operation in the \yorktown device. We note that a CNOT operation within this device is composed from a sequence operations of native one- and two-qubit gates, and the reported duration is the complete duration of that sequence. The plots in Fig.~\ref{fig:yorktown_tau_temporal} include the harmonic mean of the decoherence time $T_2$ for the register pair (0,1),  the tunable duration of the CNOT gate, and the composed duty cycle metric. On approximately July 24, 2020, the $T_2$ time decreased sharply from 77~$\mu$s to 31~$\mu$s for register 0 and from 82~$\mu$s to 24~$\mu$s for register 1. A corresponding increase in the duration of the CNOT operation between registers 0 and 1, from approximately 370~ns to 441~ns, lead to an overall decrease but seemingly consistent value for the \yorktown duty cycle. These changes in device parameters led to a sharp decrease in the duty cycle from $107.2$ to $30.9$ as seen in Fig.~\ref{fig:yorktown_tau_temporal}.
\par 
Figure~\ref{fig:yorktown_tau_temporal} also present the Hellinger distance for the duty cycle, calculated using histograms based on a running series of 3-month data. The dash redline indicates the mean for the plotted series and highlights the fluctuations in the distance. However, it is notable that the distance decreases sharply (indicating stability of the metric) late in the series due to our use of a three-month moving average for the Hellinger distance.

\subsection*{Addressability}
\begin{figure}[htbp]
\begin{center}
\includegraphics[clip,width=\columnwidth]{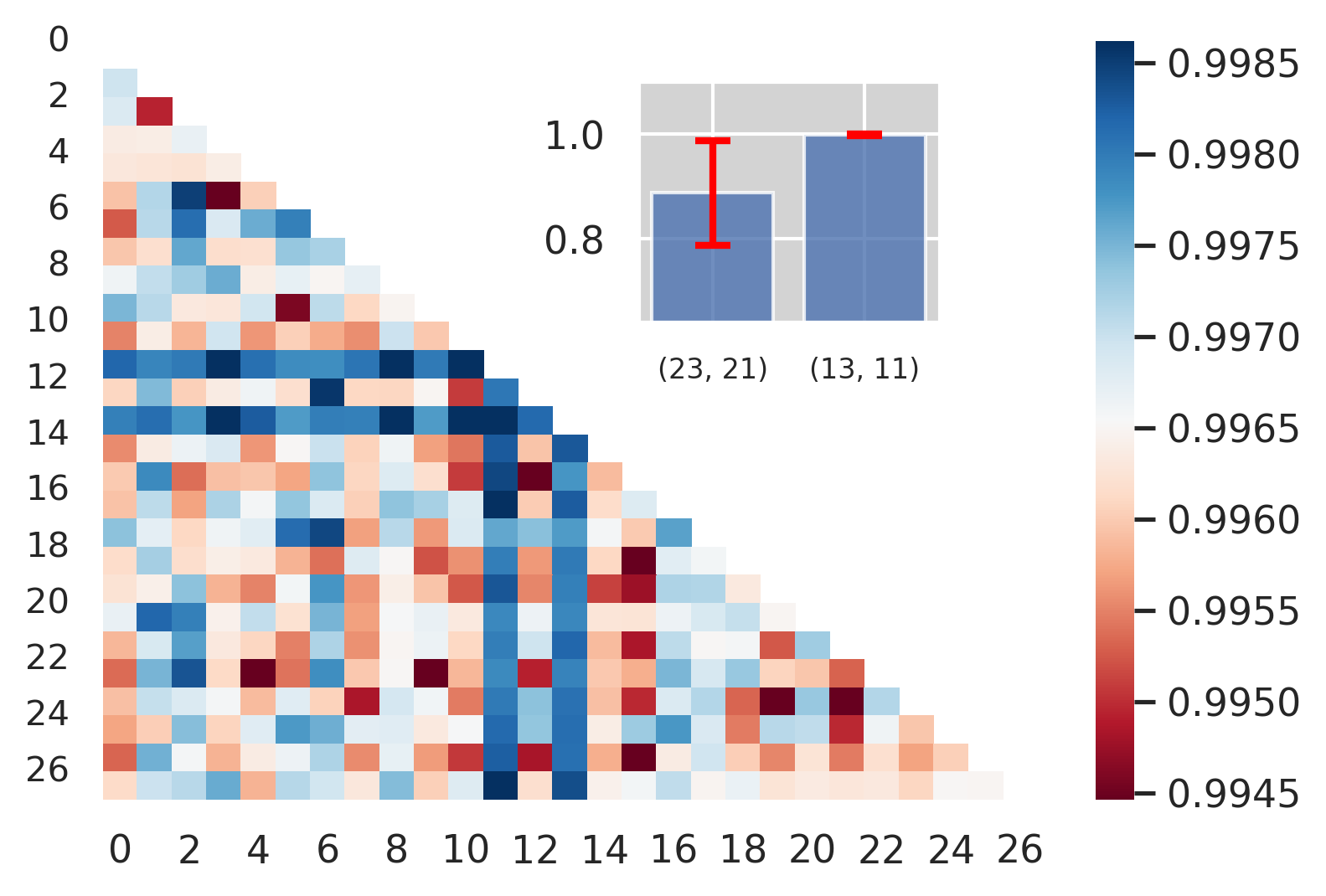}%
\caption{
\protect Addressability of register pairs in the \toronto device sampled 08:00-08:30 AM (UTC-5) on 11 December 2020. This data corresponds to the register prepared in the separable fiducial state. The inset shows the range i.e. the lowest and highest values for addressability. The average of (23,21) is the lowest value at 0.887 while all other values lie in the range [0.992, 1). The outlier is the only value that does not appear in the plot.
}
\label{fig:addr}
\end{center}
\end{figure}

Addressability quantifies how well individual register measurements are differentiated, and we quantify these correlations between individual pairs of measurements using the normalized mutual information $\eta$ and associated fidelity $F_A$ introduced in Eq.~(\ref{eq:FM}). Figure~\ref{fig:addr} plots the addressability $F_A$ of the \toronto device when tested by encoding a fiducial separable state $\ket{0,0}$ in each register pair. Whereas  ideal performance would yield $F_A = 1$ for each register pair, the heatmap highlights how the addressability $F_A$ between register pairs varies across the entire 27-element register.  
\par
The inset compares the limits of this behavior by showing the lowest valued addressability of 0.8875 for register pair (23, 21) and the highest value of 0.9989 for register pair (11, 13). We highlight that the latter pair of registers also yields the largest distance between initialization fidelity shown in Fig.~\ref{fig:toronto_FI_spatial_birch}. 
\par
A similar analysis of the addressability may be performed by first encoding a Bell-state within the register pair. As a maximally entangled state, the ideal addressabilty would $F_A = 0$ as the measurement outcomes should be strictly correlated. We limited this analysis to only nearest-neighbor register pairs as defined by the hardware connectivity, cf. Fig.~\ref{fig:toronto}, in order to minimize the number of CNOT operations used in preparing the entangled state. The resulting addressability demonstrates similar variability across register pairs.

\section*{Discussion}
\label{conc}
Fluctuations in device parameters represent a significant concern for the reproducibility of NISQ computing demonstrations. Many current experimental demonstrations rely on quantum circuits calibrated immediately prior to program execution and tuned during runtime. While this approach is successful for singular demonstrations, the resulting circuits and calibrations are implicitly dependent on the device parameters, which fluctuate significantly over time, space, and technology.
\par
Our demonstration of a framework to evaluate the stability of NISQ devices monitors multiple metrics that characterize stability. By connecting with fundamental criteria for quantum computing, the metrics in Table~\ref{tab:dvz} characterize basic quantum operations that interrogate the temporal and spatial stability of a NISQ device. We have evaluated the stability of several experimental devices by defining statistics that measure changes in device metrics. 
\par
Our contribution has identified the stability of NISQ devices as a feature of fundamental importance to current testing and evaluation. These analyses are readily extended to additional characteristics as well as metadata analysis of the calibration date, optimal pulse data, and other features about device operations.
\par
The long-term reproducibility of results from experimental quantum computer science, and in particular those using NISQ devices, hinges on the stable and reliable performance of the computer device. Without additional efforts to make current experimental results reproducible across either time or other devices, the knowledge and insights gained from today's burgeoning field of quantum computer research may be undercut by low confidence in the reported results. Our results indicate clearly distinguishable differences in the Hellinger distance for several metrics.

\section*{Acknowledgments}
This work is supported by the Department of Energy (DOE) Office of Science, Early Career Research Program. This research used computing resources of the Oak Ridge Leadership Computing Facility, which is a DOE Office of Science User Facility supported under Contract DE-AC05-00OR22725. SD thanks Peter Lockwood for support. A preliminary version of this work was published in the IEEE Quantum Computing Conference 2020 \cite{dasgupta2020characterizing}.

\section*{Author information}
These authors contributed equally: S.~Dasgupta, T.~S.~Humble

\bibliographystyle{unsrt}
\bibliography{references.bib}

\newpage
\clearpage 
\renewcommand{\thefigure}{A\arabic{figure}}   
\setcounter{figure}{0}
\setcounter{section}{1}
\renewcommand{\thesection}{\Alph{section}} 
\section*{Appendix}\label{sec:Appendix}
This appendix includes details on the data collection methods, the analysis methods and supplementary results from this analysis.
\par
\subsection{Data Collection}
For the $n=5$ \yorktown device, we use calibration data from the publicly available data set provided by the IBM Quantum Experience \cite{ibm_quantum_experience_website}. This data set includes daily calibration data collected for the \yorktown device over a period of 22 months from March 2019 to December 2020. This data set includes a total of 673 calibration records. The first record has a time stamp of 2019-02-27 06:56:32-05:00 and the last record has a time stamp of 2020-12-30 01:18:20-05:00. We list the available fields in each record below. We note a change in the records that occurred in September 2019 when the gate duration data became available. This yielded 445 records with gate duration data. Otherwise, each calibration record contains the following data identified by the indicated fields:

\begin{enumerate}
\item Date and time when the device properties were last updated as last\_update\_date

\item Register readout error estimates as, e.g. q0\_readout\_err

\item Date and time when a register readout was calibrated as, e.g. q0\_readout\_err\_cal\_time

\item CNOT gate error estimates as, e.g. cx01\_gate\_err

\item Date and time when a CNOT gate error measurement was calibrated as, e.g. cx01\_gate\_err\_cal\_time

\item Register decoherence time estimate, the unit of measure, and the date and time of the characterization as, e.g. q0\_T2, q0\_T2\_unit, and q0\_T2\_cal\_time

\item CNOT gate length and the unit of measure as, e.g. cx01\_gate\_length, cx01\_gate\_length\_unit

\item Date and time when a CNOT gate length was calibrated as, e.g. cx01\_gate\_length\_cal\_time

\end{enumerate}

\par
For the $n=27$ \toronto device, we generated a second data set on 11 December 2020 that spans the periods of time 8:00-8:30am, 11:00-11:30am, 2:00-2:30pm, 5:00-5:30pm, 8:00-8:30pm, and 11:00-11:30pm (UTC-5). The collected data included the readout of all $n=27$ register elements when the initial state is prepared to be in the all-zeros computational basis state. The total number of observations in this data set was $5,750,784$ register results. Each register element was sampled for  $212,992$ sequential data points. We divided this temporal set into contiguous sets of 1000 data points each, e.g., the first set contains the first $1000$ observations for each of the 27 registers as a $1000\times 27$ binary matrix. Set 1 contains the next set of $1000$ observations for each of the 27 qubits - another $1000\times 27$ binary matrix, etc. For each data set, we calculated the addressability $F_A$ for each of the possible 351 register pairs and each pair was characterized by 211 observations for $F_A$.
\par 
A similarly structured data set was constructed from the readout of all $n=27$ register elements when the initial state of register pairs are prepared in the Bell-state triplet state. As above, we collected data from \toronto following a canonical Bell-state preparation circuits applied to nearest-neighbor register locations. There are 28 such nearest-neighbor pairs in the \toronto layout. We characterized the measurement outcomes of a Bell-state preparation circuit for each of these 28 pairs on December 11, 2020 between 11pm-11:30pm. Each experiment was recorded with 8192 observations.
\par 
We generated a third data set composed from data collected across a family of different IBM devices. These devices are labeled as yorktown, bogota, rochester, paris, and athens. The calibration records collected from the IBM Quantum Experience were used to construct a data set for each device consisting of the update time, the readout errors, the readout calibration time, the CNOT gate error rates and the CNOT gate calibration time. This was collected over the period of 27 February 2019 to 31 December 2020.
\par


\subsection{Stability Analysis}
For analysing device stability, we compute the distance between empirical distributions. When calculating the Hellinger distance between histograms of registers at different layout locations, the full time series available was always used. When performing the inter-device stability analysis, the Hellinger distance was measured between the histograms generated by the full time-series for each device. For the temporal stability analysis, each data point represents the distance for observed histograms spaced 1 month apart, in which a sliding window was used. Each histogram bins the data collected over a 3-month period, such that there are 90 daily data values in each histogram.
\par
We monitor temporal stability by tracking temporal changes in the Hellinger distance between distributions at different times as
\begin{equation}
\mathcal{H}(t) = d_{H}(\mathcal{F}_X(t-\tau), \mathcal{F}_X(t))
\end{equation}
where $\mathcal{F}_{X}(t)$ is the distribution of metric $X$ at time $t$ and $\tau$ is a reference duration. For temporal stability analysis of the initialization fidelity , we set $\tau = 1$ month to reveal changes over a moving 1-month window. For the temporal stability analysis of the gate fidelity, we track $\mathcal{H}(t)$ with respect to a reference time $t_0$ such that $\tau = t-t_0$ to reveal variations relative to when an experiment was first conducted ($t_0$). 
\par 
We monitor spatial stability by tracking spatial changes in the Hellinger distance between distributions at different register locations as
\begin{equation}
\mathcal{H}(i,j) = d_{H}(\mathcal{F}_X(i), \mathcal{F}_X(j))
\end{equation}
where $\mathcal{F}_X(i)$ is the corresponding distribution of metric $X$ at register $i$ over all available temporal data. Similarly, we monitor the stability of a metric between different devices by using register locations $i$ and $j$ that correspond to locations in different devices. 
\subsection{Supplemental Figures}
\begin{figure}[htbp]
\begin{center}
\includegraphics[width=\columnwidth]{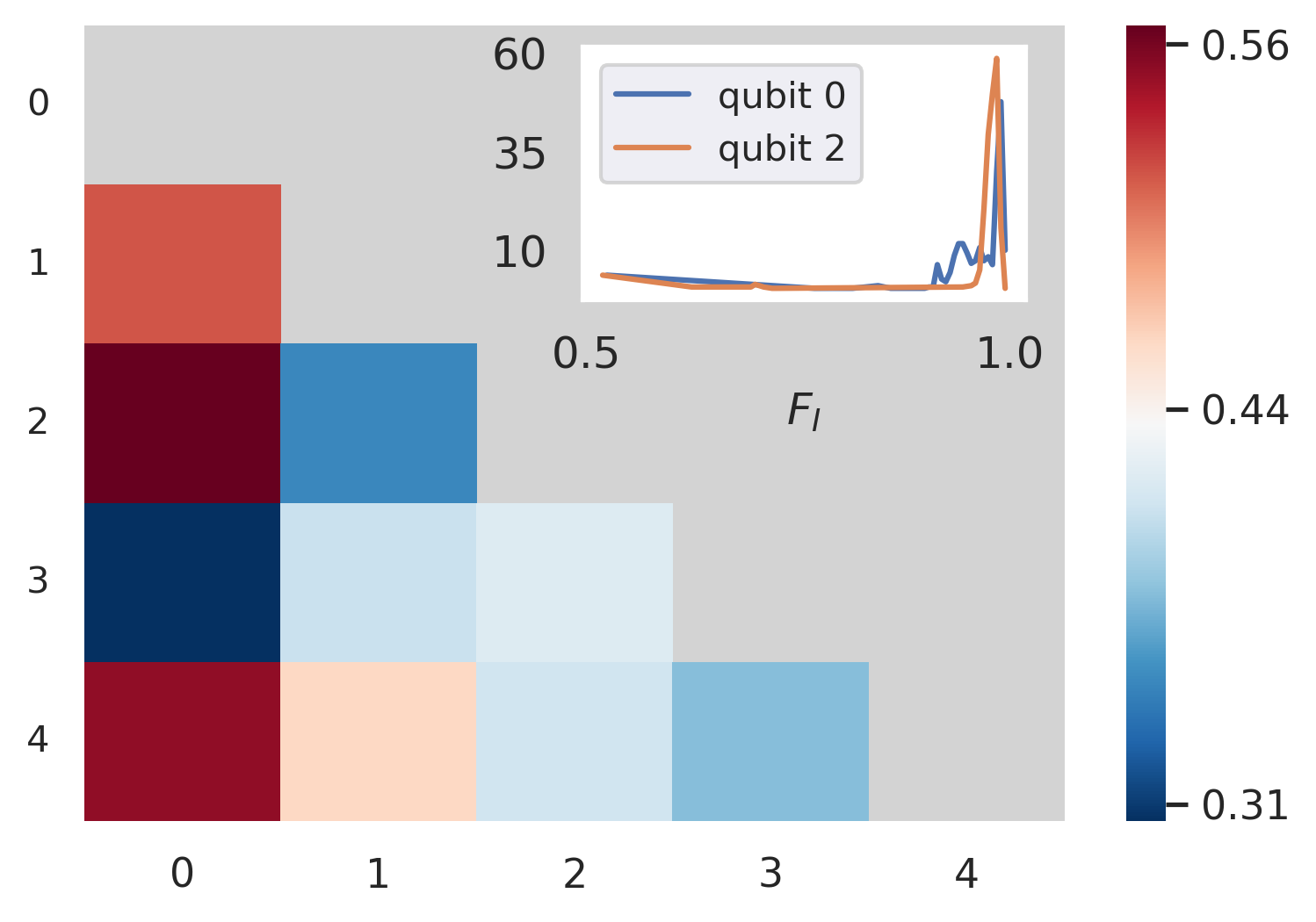}%
\caption{
\protect Spatial stability of the initialization fidelity for the \yorktown device from May 2019 to December 2020, where the inset highlights the registers with the maximum distance.
}
\label{fig:yorktown_FI_spatial_birch}
\end{center}
\end{figure}

Figure~\ref{fig:yorktown_FI_spatial_birch} plots the spatial stability analysis of initialize fidelity for the $n=5$ \yorktown device from the first data set. The data demonstrates variation in the distance between register pairs that range from approximately 0.35 to as larger as 0.55. The inset shows the distributions of $F_I$ for the largest distance observed.
\par
\begin{figure}[htbp]
\begin{center}
\includegraphics[clip,width=\columnwidth]{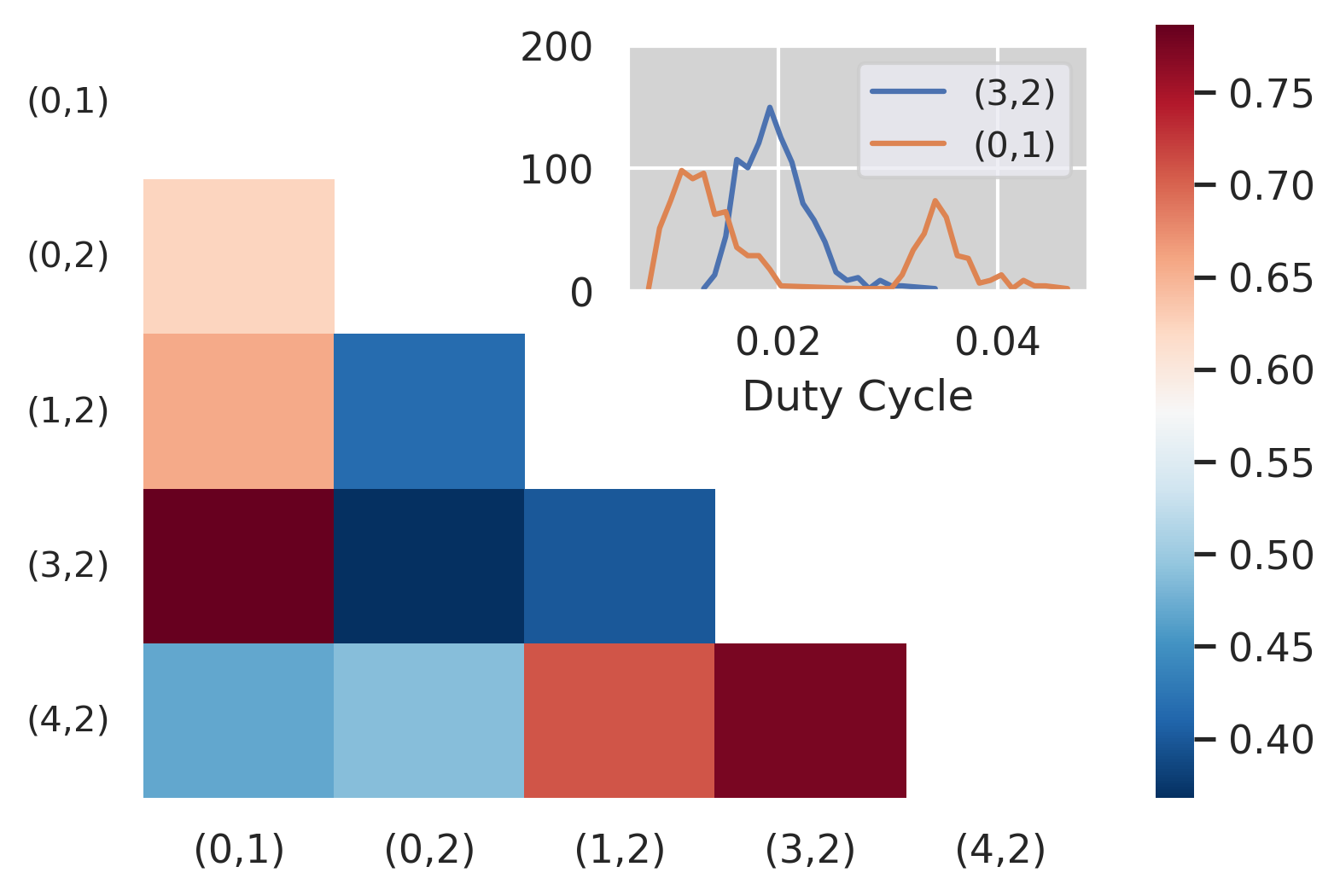}
\caption{
\protect The spatial stability of the duty cycle $\tau$ for yorktown. The inset shows the experimental histograms for register pairs (0,1) and (2,3) which are separated by the largest Helligner distance of 0.789.}
\label{fig:yorktown_tau_spatial_spruce}
\end{center}
\end{figure}

Figure~\ref{fig:yorktown_tau_spatial_spruce} plots the spatial stability of the duty cycle for the $n=5$ \yorktown device from the first data set. The data demonstrates variation in the distance between register pairs that range from approximately 0.40 to as larger as 0.789. The inset shows the distributions of the duty cycle for the largest distance observed. Distributions for register pairs (0,1) and (3,2) showed a large distance of 0.789. The spatial stability pattern for the duty cycle does not match the pattern observed for the initialization fidelity in Fig.~\ref{fig:yorktown_FG_spatial_fir}, hinting at more different sources of errors for these transient behaviors.
\par
\begin{figure}[htbp]
\begin{center}
\includegraphics[clip,width=\columnwidth]{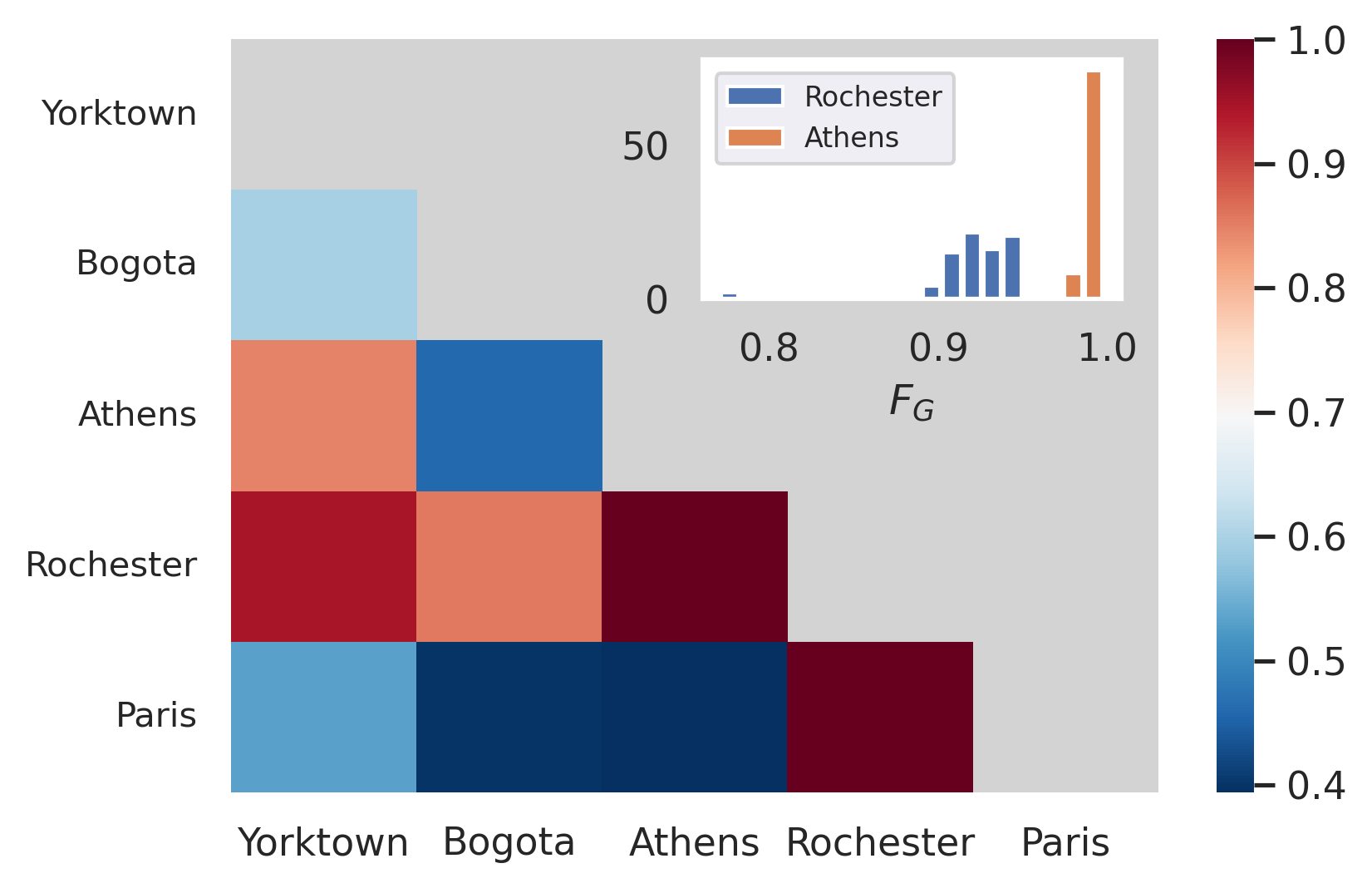}
\caption{Inter-device stability of the gate fidelity $F_G$ for the CNOT gate for a family of five devices for the time period Mar 2019 - Dec 2020. The inset shows the distribution of $F_G$ for the CNOT gate from the \athens and \rochester devices which have the highest corresponding Hellinger distance of nearly 1.}
\label{fig:pine}
\end{center}
\end{figure}

Figure~\ref{fig:pine} plots the spatial stability of gate fidelity varies across the five devices from the third data set. The figure plots a pair-wise heatmap based on the Hellinger distance between gate fidelity distributions for these devices. The inset shows an example of the underlying distributions between the \rochester and \athens devices. These two devices demonstrate significant differences in distribution as indicated by a Hellinger distance of 1.0 (no overlap).
\par
\begin{figure}[htbp]
\begin{center}
\includegraphics[clip,width=\columnwidth]{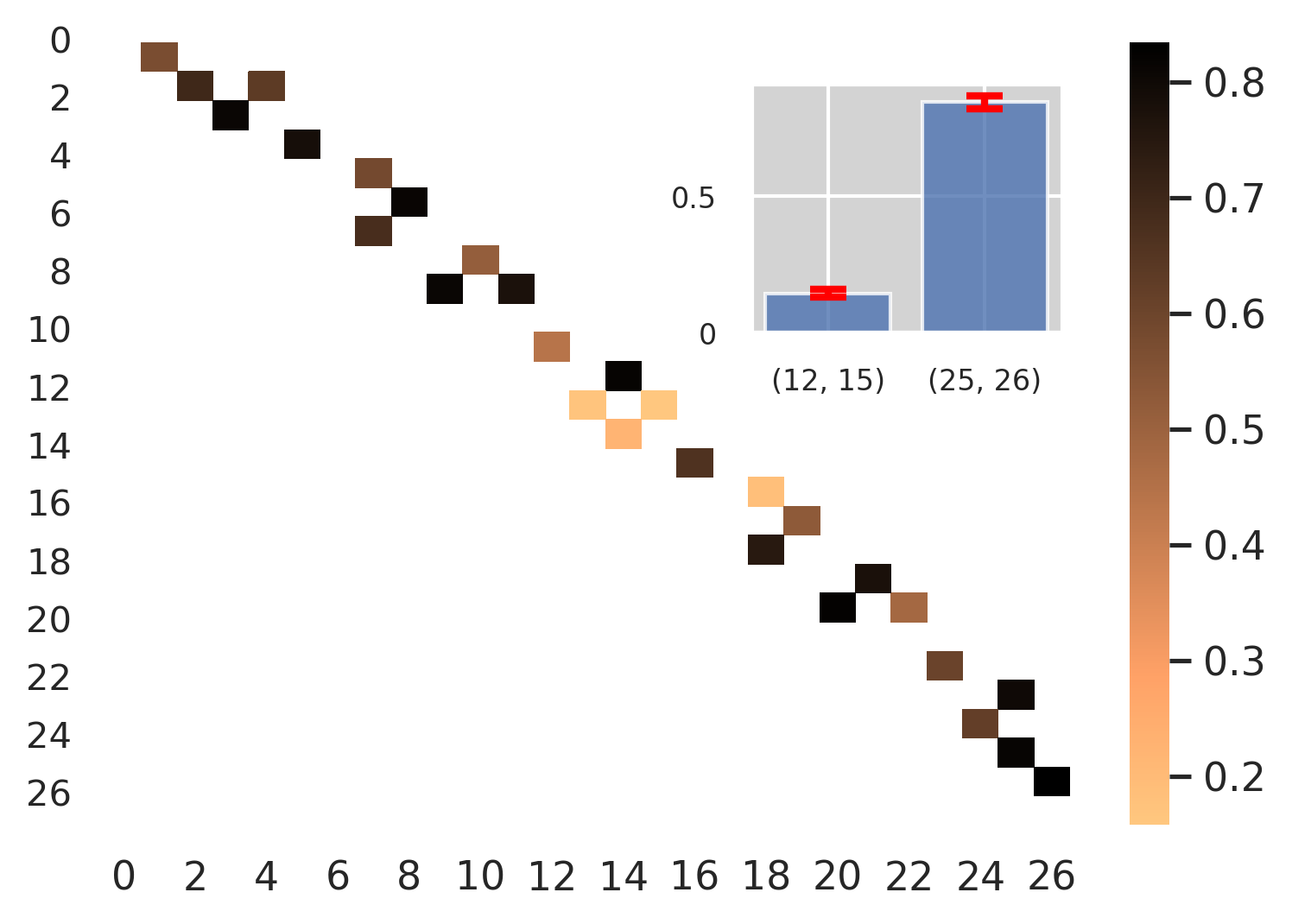}%
\caption{\protect Normalized mutual information of register pairs in the \toronto device sampled 11:00-11:30 PM (UTC-5) on 11 December 2020. Data corresponds to register pairs prepared in the Bell state and the inset shows the range of the lowest and highest values for the normalized mutual information.}
\label{fig:bell_spatial_nmi_cotton}
\end{center}
\end{figure}

 
Figure~\ref{fig:bell_spatial_nmi_cotton} plots the normalized mutual information (NMI) for Bell-state encoded states. There only 28 register pairs that support direct preparation of a Bell state and the figure plots the NMI for those pairs that support these states. For this state, strictly correlated measurement outcomes are expected and the NMI should be maximal. The inset shows an example of the range of the observed NMI between the Bell pairs. We find the NMI between registers 12 and 15 was least with a mean of $\eta = 0.14\pm 0.014$, while the largest NMI was observed for registers 25 and 26 with a mean of $\eta= 0.84 \pm 0.023$. 
\par 


\end{document}